\documentclass[reprint,amsmath,amssymb,aps,nofootinbib]{revtex4-2}
\usepackage{graphicx}
\graphicspath{ {images/} }
\usepackage{epstopdf}
\usepackage{hyperref}

\usepackage{color}
\definecolor{green}{rgb}{0,0.5,0}
\usepackage{xcolor}
\usepackage{textcomp}

\begin{document}

\title{Granular gases under resetting}
\author{Anna S. Bodrova$^1$, Aleksei V. Chechkin$^2$$^3$$^4$,
Awadhesh Kumar Dubey$^5$}

\address{$^1$ Moscow Institute of Electronics and Mathematics, HSE University, 123458, Moscow, Russia}

%\affiliation{Moscow Institute of Electronics and Mathematics, HSE University, 123458, Moscow, Russia}
%\author{Aleksei V. Chechkin}

\address{$^2$Institute of Physics and Astronomy, University of Potsdam, 14476 Potsdam, Germany}

\address{$^3$Faculty of Pure and Applied Mathematics, Wroclaw University of Science and Technology, Wyspianskiego 27, Wrocław, 50-370, Poland}

\address{$^4$Max Planck Institute of Microstructure Physics, Weinberg 2, 06120 Halle, Germany}

\address{$^5$Department of Pure and Applied Physics,
Guru Ghasidas Vishwavidyalaya,
Koni, Bilaspur-495009
Chhattisgarh, India.}

\date{\today}

\begin{abstract}
We investigate the granular temperatures in force-free granular gases under exponential resetting. When a resetting event occurs, the granular temperature attains its initial value, whereas it decreases because of the inelastic collisions between the resetting events. We develop a theory and perform computer simulations for granular gas cooling in the presence of Poissonian resetting events.
%for the dependence of the mean granular temperature and its variance on the resetting rate. 
We also investigate the probability density function to quantify the distribution of granular temperatures. Our theory may help us to understand the behavior of nonperiodically driven granular systems.
\end{abstract}

\maketitle

\section{Introduction}

A process with resetting breaks at a certain point and starts anew. 
%A process with resetting gets interrupted at certain points and starts anew. 
Resetting has been widely studied recently and has numerous applications \cite{reviewres, reviewresgupta}. 
This significantly accelerates search processes \cite{SokChech,shlomimfpt,shlomi} and improves the efficiency of computer algorithms \cite{computerscience,lor}. The resetting processes have been observed in various fields. In biology, resetting occurs in enzyme-catalyzed reactions, described in terms of Michaelis-Menten kinetics \cite{chemistry,bio1,bio2}, transcription \cite{rna}, and mobility of animals \cite{animals}. In economic society models, resetting may represent a loss of wealth due to catastrophic events \cite{ecotax}. Geometric Brownian motion with stochastic resetting can be used to describe income dynamics \cite{ecoin, gbmresralf}. Resetting may represent eye movements while viewing and recognizing the patterns \cite{psi1,psi2}.

In the first study, resetting was considered for particles exhibiting overdamped Brownian motion \cite{PRLinitial,PREmaj}. Subsequently, many other processes with resetting have been investigated \cite{reviewres, reviewresgupta, TS}, such as underdamped Brownian motion \cite{underdamped, underdamped24} the Ornstein-Uhlenbeck process \cite{OU2015, OU2023}, continuous-time random walks \cite{Annactrw,MV2013,MC2016,Sh2017,ctrw,ctrwres,ctrwres1}, L\'evy flights \cite{levy1,levy2}, L\'evy walks \cite{china}, heterogeneous diffusion processes \cite{hetero, fbm}, fractional Brownian motion \cite{fbm}, geometric Brownian motion \cite{gbm,gbm1}, scaled Brownian motion \cite{Annare,Annanonre}, and resetting on networks \cite{net,net2}. Initially, the resetting process was considered as a jump to the starting point. Later, other types of return processes were considered, such as partial resetting \cite{partial, partial1, partial2}, return at constant velocity \cite{shlomi, shlomi1,shlomi2,campos,Annasmooth,Annasmooth2d, radice}, and under the action of an external potential \cite{pot1, pot2, poti}. The latter phenomenon was observed in experiments \cite{exp1, exp2, exp3}. Most investigations have dealt with the resetting of coordinates. However, the resetting of other quantities has also been considered, such as the diffusion coefficient in scaled Brownian motion \cite{Annare,Annanonre} and velocity in underdamped Brownian motion \cite{underdamped, underdamped24}, 

%Here we could add citations of other papers and the corresponding discussion from underdamped24 about resetting of the velocity!
%Gr2.jpg
%\begin{figure}\centerline{\includegraphics[width=1.0\columnwidth]{Gmd.jpg}}\caption{A typical evolution of granular temperature $T(t)$ in granular systems with exponential resetting obtained in event-driven simulations. The resetting rate $r=0.1$, the characteristic time of the granular temperature decay $\tau_0=15.3$, the number of particles $N_{\rm par}=10^5$. Between the resetting events the granular temperature decays according to Haff's law (Eq.~\ref{Thaff}).} \label{Gr2}\end{figure}

In this study, we investigate the resetting in granular systems. There are numerous examples of granular systems, such as stones and sand in the building industry; grains, sugar, salt, and cereals in the food industry; and different kinds of powders in chemical and cosmetic production \cite{PhysGranMed, SanPowGr, DryGranMed, GranRev}. Granular materials consist of macroscopic solid particles but can also exhibit behavior similar to that of conventional phases of matter. They can flow like liquids down an incline and be in a gaseous state, when the particles move at high velocities, and the typical distance between granular particles exceeds their size. 

Numerous examples of granular gases include dust devils, large interstellar dust clouds \cite{inter}, protoplanetary discs, planetary rings \cite{ringbook,rings, pnas}, and asteroids \cite{aster}. There are several different possibilities for simulating granular systems in computer experiments \cite{compbook}, such as Monte Carlo (MC) and molecular dynamics (MD). The former allows for more rapid simulations of large systems, whereas the latter resembles the properties of real systems.

Granular gases have many similarities with molecular gases \cite{GranRev, book, mehta}. However, the major difference between them is the dissipative collisions, where part of the kinetic energy of the granular gases is lost \cite{book}.  The motion of the particles slows down, and the system homogeneously cools according to Haff's law \cite{haff}. %the system evolves force-free by itself and cools homogeneously according to Haff's law \cite{haff}. 
Although clustering and vortex formation can be observed at the later stages of the evolution of granular systems \cite{book}, the homogeneous state may persist long enough even for a polydisperse system \cite{zippelius}. 
Although granular systems may be highly polydisperse \cite{anna2024}, we restrict our consideration to three-dimensional unicomponent granular gases.

%Granular gases represent diluted granular systems \cite{GranRev, book, mehta} in which the typical distance between their components significantly exceeds their dimensions. 

 %We apply both methods in our current study.

%In the homogeneous cooling state,  granular gases remain force-free and lose their kinetic energy during collisions \cite{book}. To compensate for  dissipation, energy should be inserted into the system. 
To maintain a gaseous state, energy should be supplied to the system. There are different possibilities for providing energy supply to granular systems: vibrating \cite{vib1, vib2} or rotating \cite{rot} walls, external electrostatic \cite{el} or magnetic forces \cite{magn1, magn2, magnsperl, magnsperl21}.  

The driving of granular gases is often described in theory and computer simulations in terms of uniform heating \cite{therm, caftherm, santherm, prasadtherm} or heating through the boundaries \cite{zon1, zon2}. 
The heating events may also occur as rare but powerful energy injections into a single randomly selected particle \cite{kangextreme,bennaim}. 
In addition, particles may be reenergized such that their velocities are drawn from a Maxwellian distribution with a typical energy proportional to the system size \cite{bennaim}. Velocity distribution functions have been studied for various driving mechanisms. It has been shown that the velocity distribution of granular gases is governed by the heating-dissipation rate, which is given as the ratio between the average number of heating events and the average number of collisions \cite{zon1, zon2}. However, the quantitative dependence of the most important macroscopic parameters, such as granular temperature, on the ratio between the average number of heating events and average number of collisions has not been derived before. This dependence may be universal in systems with different driving mechanisms.

%This we should check and discuss after the simulations are done

We describe rare heating events within the framework of resetting of granular temperatures. 
We consider Poissonian resetting, which means that the resetting event may occur with equal probability at any given time. During the instantaneous resetting event, the granular temperature attains its initial value, $T_0$. Between the resetting events, the system evolves force-free by itself and cools homogeneously according to Haff's law \cite{haff}. 

We proceed as follows. In Section II, we briefly review the evolution of the force-free granular gases. In Section III, we present our theory of the Poissonian resetting of  granular gases and study the probability density function (PDF) of the granular temperatures and its moments. We discuss event-driven numerical simulations and compare them with theory in Sec. IV. Finally, we present our conclusions in Sec. V.

\begin{figure}\centerline{\includegraphics[width=1.0\columnwidth]{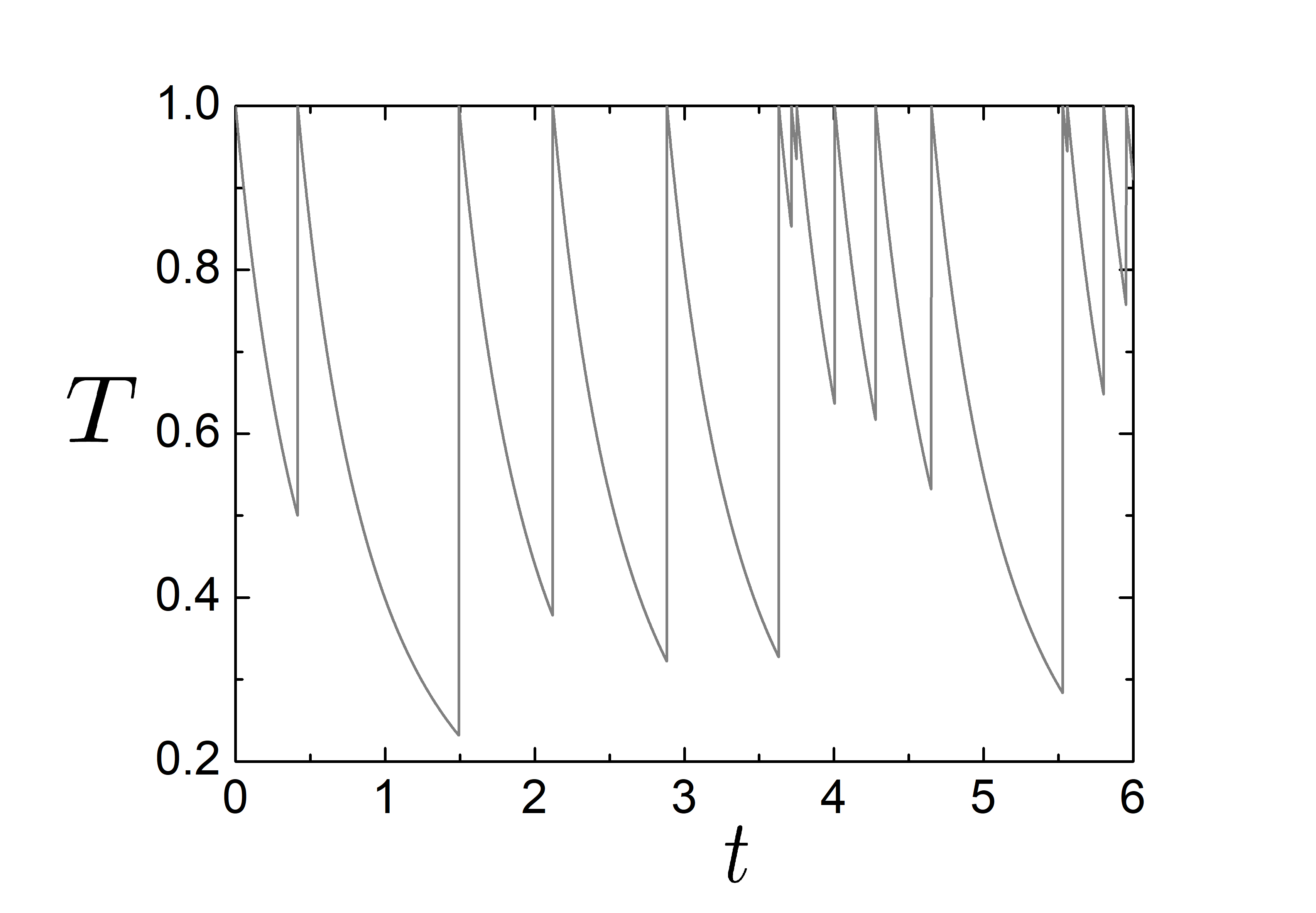}}\caption{A typical evolution of granular temperature $T(t)$ in granular systems with exponential resetting. The resetting rate $r=2$, the characteristic time of the granular temperature decay $\tau_0=1$, the initial granular temperature $T_0=1$.} \label{Gr2}
\end{figure}

\section{Haff's law}

The granular temperature is one of the most important parameters in the description of granular gases. This is defined in terms of the mean kinetic energy \cite{book}
\begin{equation}
\frac32 n T(t)=\frac{m\langle v^2\rangle}{2} =\int d {\bf v} f\left({\bf v}, t\right)\frac{m v^2}{2}\,.
\end{equation}
Here, $m$ is the mass of the granular particle, $v$ corresponds to its velocity, $n$ is the number density of particles in the granular gas, and $f\left({\bf v}, t\right)$ is the velocity distribution function, which is assumed to be Maxwellian for simplicity, despite the slight deviations from the Maxwellian form obtained in both theory \cite{gold,vane,huth,brilpo2000,bripo,annaphysa,annamsu,annapre} and experiments \cite{Sperl}. Owing to the dissipative collisions, the granular temperature in the granular gas gradually decreases. The evolution of granular temperature obeys the following differential equation \cite{book}:
\begin{equation}\label{difftemp}
\frac{dT(t)}{dt}=-T(t)\xi(t)\,.
\end{equation}
Here the cooling rate $\xi(t)$ is equal to \cite{book}
\begin{equation}\label{xit}
\xi(t) = \frac{4}{3}\left(1-\varepsilon^2\right)n\sigma^2\sqrt{\frac{\pi T}{m}}\,,
\end{equation}
where, $\sigma$ is the diameter of the granular particles, and the restitution coefficient $\varepsilon$ quantifies the dissipative losses during the collision of granular particles as follows \cite{book,GranRev}:
\begin{equation}
\label{rc} \varepsilon = \left|\frac{\left({\bf v}^{\,\prime}_{ki} \cdot {\bf e}\right)}{\left({\bf v}_{ki} \cdot {\bf
e}\right)}\right| \, .
\end{equation}
Here,  ${\bf v}_{ki}={\bf v}_{k}-{\bf
v}_{i}$ and ${\bf v}^{\,\prime}_{ki}={\bf v}_{k}^{\,\prime}-{\bf v}_{i}^{\,\prime}$ are the relative velocities before and after a  collision, respectively, and ${\bf e}$ is a  unit vector directed along the inter-center vector  at the collision instant.  For simplicity, the restitution coefficient is often assumed constant \cite{book}. It is easy to implement in analytical calculations, and can be considered a basic reference model. The post-collision velocities  ${\bf v}_{k}^{\,\prime}$ and ${\bf v}_{i}^{\,\prime}$ are  related to the pre-collision velocities ${\bf v}_{k}$ and ${\bf v}_{i}$ as follows \cite{book}:
\begin{equation}\label{v1v2} {\bf v}_{k/i}^{\,\prime} = {\bf v}_{k/i} \mp  \frac{1+\varepsilon}{2}({\bf v}_{ki} \cdot {\bf e}){\bf e} \, .\end{equation}
Differential equation (Eq.~\ref{difftemp}) with the cooling rate (Eq.~\ref{xit}) can be solved explicitly, and the temperature obeys Haff's law \cite{haff}
\begin{equation}\label{Thaff}
T(t) = T_0\left(1+\frac{t}{\tau_0}\right)^{-2}\,,
\end{equation}
where $T_0=T(0)$ is the initial granular temperature. The inverse temperature relaxation time is equal to half the cooling rate (Eq.~\ref{xit}) at the initial time: $\tau_0^{-1}=\xi(0)/2$.
%\begin{equation}\label{tau0}\tau_0^{-1}=\xi(0)/2\,.\end{equation}
It may be expressed also in terms of the mean collision time $\tau_c(t)$ in the following way:
\begin{equation}\label{tau02}
\tau_0^{-1} = \frac{1-\varepsilon^2}{6}\tau_c^{-1}(0)
\end{equation}
The mean collision time may be derived as \cite{book}
\begin{equation}\label{tauc}
\tau_c^{-1}(t)=4n\sigma^2g_2(\sigma)\sqrt{\frac{\pi T}{m}}
\end{equation}

\section{Theory}
\subsection{Poissonian resetting}

We assume that resetting events occur according to Poisson process \cite{poi} with a constant rate $r$. Thus, the average time between resetting events is equal to $1/r$. The probability $r dt$ for the resetting event to occur during the time interval $\left(t, t + dt\right)$ does not depend on time $t$ and depends only on the time interval $dt$. The waiting time distribution between resetting events is
\begin{equation}
\psi (t) = r{e^{ - rt}}\,.
\label{pdfexp}
\end{equation}
The survival probability $\Psi(t)$ is defined as the probability that no resetting event occurs between zero and $t$,
\begin{equation}
\Psi (t) = 1 - \int\limits_0^t {\psi (t')dt'}  =  {e^{ - rt}}\,.
\label{surv}
\end{equation}

We perform computer simulations using both event-driven MD simulations (described in Section IV) and simple MC simulations. In the latter, the granular temperature evolves initially  according to Haff's law (Eq.~\ref{Thaff}). If the resetting event takes place at a constant rate $r$, the granular temperature takes the initial value $T_0=1$. Subsequently, it starts to decrease again. The typical evolution of a granular system with a resetting rate $r=2$ is shown in Fig.~\ref{Gr2}.

\begin{figure}\centerline{\includegraphics[width=0.98\columnwidth]{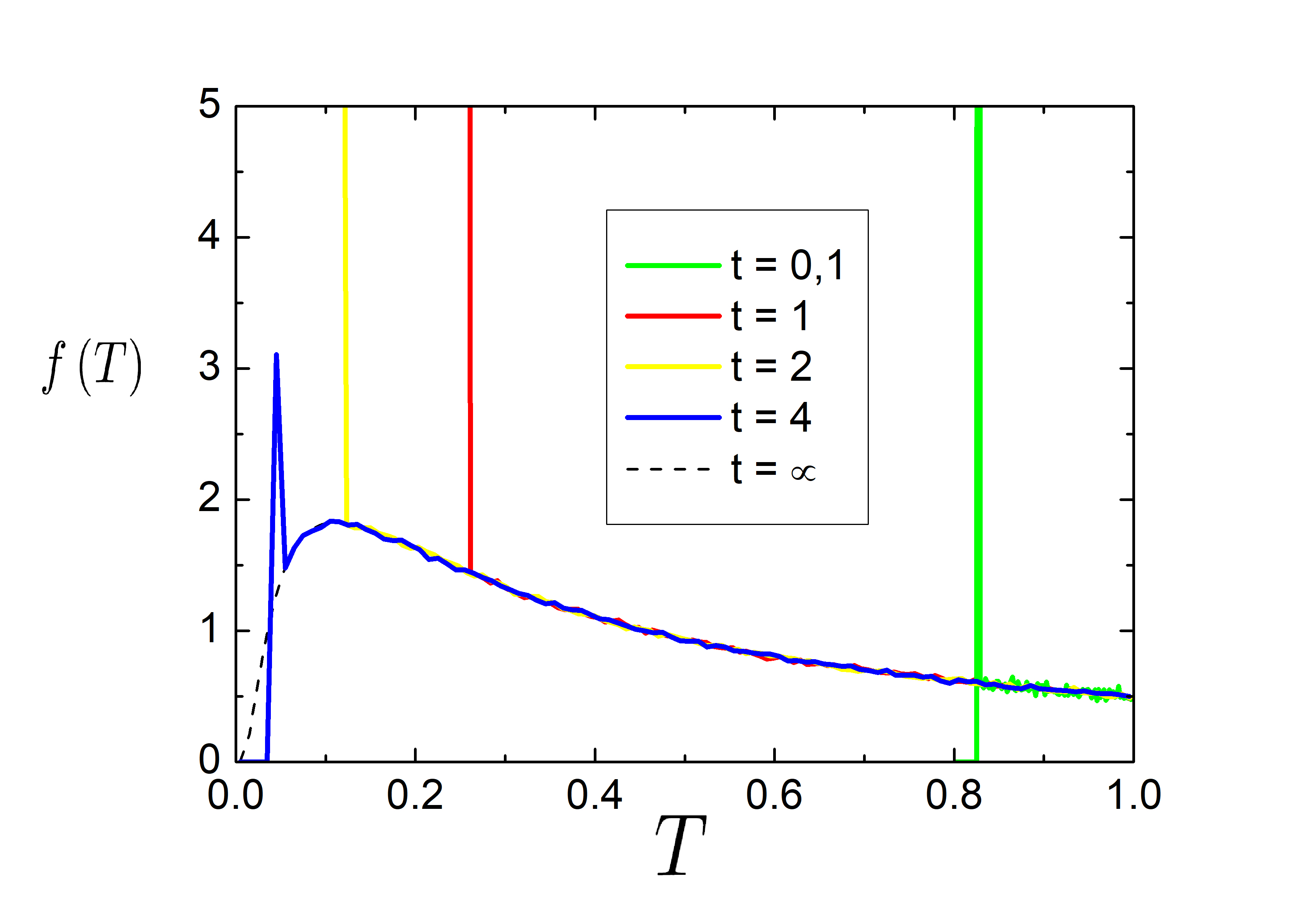}}\caption{Evolution of the probability density function of granular temperatures $f(T,t)$ (Eq.~\ref{ftv}), $\tilde{r}=1$. The peaks may be obtained by introducing the corresponding times into the Haff's law (Eq.~\ref{Thaff}). Solid lines correspond to results of MC simulations, dashed line - to theory at long times, $t\to\infty$ (Eq.~\ref{fT0}). 
} \label{GfTmany}
\end{figure}

\subsection{Probability density function}

Let us consider evolution of a granular temperature with an arbitrary power-law constant $a$
\begin{equation}\label{Haffgen}
T(t) = T_0\left(1+\frac{t}{\tau_0}\right)^{-a}\,.
\end{equation}
The power-law constant $a$ may vary for different granular cooling scenarios, for example in the presence of clustering \cite{puri16}. 
For the viscoelastic granular gas model $a=5/3$ \cite{book,brilpo2000,annamsu}.

The probability density function (PDF) $f_0(T)$ of the granular temperature distribution without resetting has the form of a Dirac delta-function:
\begin{eqnarray}\label{f0delta}
f_0(T,t)= \delta\left(T-T_0\left(1+\frac{t}{\tau_0}\right)^{-a}\right)
\end{eqnarray}
The probability density function (PDF) with resetting can be obtained according to \cite{reviewres}:
\begin{eqnarray}
f\left(T,t\right)=  f_0(T,t) e^{-rt}+r\int\limits_0^{t}  f_0(T,\tau) e^{-r\tau} d\tau\,.\label{intint}
\end{eqnarray}
%\begin{eqnarray}\nonumber&&f\left(T,t\right)=  \delta\left(T-T_0\left(1+\frac{t}{\tau_0}\right)^{-a}\right)\!e^{-rt}+ \\&&+r\int\limits_0^{t}  \delta\left(T-T_0\left(1+\frac{\tau}{\tau_0}\right)^{-a}\right)\!e^{-r\tau} d\tau\,.\end{eqnarray}
The first term accounts for the realizations where no resetting occurs up to observation time $t$. In this case the evolution of the granular temperature is determined by the generalized form of Haff's law (Eq.~\ref{Haffgen}), and the PDF takes the form of a Dirac delta-function. The second term accounts for the case in which the last resetting event occurs at time $t-\tau$,  after which no resetting occurs between $t-\tau$ and $t$ with probability $\Psi(\tau)$ (Eq.~\ref{surv}). The PDF value is determined by the time interval $\tau$ passed from the last resetting event owing to the memory loss. %, only the time elapsed since the last resetting event $\tau$ is relevant to establish the value of granular temperature at time $t$. Integration  in Eq.~(\ref{Tpsi}) yields
Introducing Eq.~(\ref{f0delta}) into Eq.~(\ref{intint}) and performing the integration, we find that the PDF has the form
%\begin{widetext}
\begin{eqnarray}\nonumber
&&f\left(T,t\right)= 
\delta\left(T-T_0\left(1+\frac{t}{\tau_0}\right)^{-a}\right)\!e^{-rt}+\frac{\tilde{r}}{aT_0}\left(\frac{T_0}{T}\right)^{1+1/a}\\
&&\!\!\times\exp\left(\tilde{r}\left(1-\left(T_0/T\right)^{1/a}\right)\right)\Theta\left(T-T_0\left(1+\frac{t}{\tau_0}\right)^{-a}\right)\nonumber\\
\label{ftv}
\end{eqnarray}
for $0<T\le T_0$ and zero otherwise. Here, we introduced the parameter $\tilde{r}=r\tau_0$ as the ratio between the characteristic temperature relaxation time and average time $1/r$ between resetting events, and $\Theta(x)$ is the Heaviside step function.
At long times $t\gg 1/r$ the probability that no resetting events have occurred tends to zero, and the first term in Eq.~(\ref{ftv}) can be neglected. By setting $t\to\infty$, we obtain the steady-state value of the PDF
\begin{equation}
f\left(T,t\right)= \frac{\tilde{r}}{aT_0}\left(\frac{T_0}{T}\right)^{1+1/a}\!\!\exp\left(\tilde{r}\left(1-\left(T_0/T\right)^{1/a}\right)\right)\,.
\end{equation}
In such a way, the PDF of reciprocal reduced granular temperatures $T_0/T$ has the form of Gamma distribution.
The mean granular temperature may be obtained using the PDF (Eq.~\ref{ftv}):\begin{eqnarray}\nonumber
&&\left\langle T(t)\right\rangle =\int_0^{T_0} f\left(T,t\right)T dT= T_0e^{-rt}\left(1+\frac{t}{\tau_0}\right)^{-a}+\\&& T_0\tilde{r}e^{\tilde{r}}\left(E_a(\tilde{r})-\left(1+\frac{t}{\tau_0}\right)^{1-a}E_a\left(\tilde{r}\left(1+\frac{t}{\tau_0}\right)\right)\right)\,.\label{Tmean}
\end{eqnarray}
with $E_a(z)$ being the generalized integro-exponential function \cite{milgram}:
\begin{equation}\label{Ea}
E_a(z)=\int_1^{\infty}\frac{e^{-zt}}{t^a}dt\,.
\end{equation}
In the long time limit $t\to\infty$ this value tends to
\begin{eqnarray}\label{Tgen}
T_c = T_0\tilde{r}e^{\tilde{r}}E_a(\tilde{r})\,.
\end{eqnarray}
The moments of granular temperatures can be obtained analogously
\begin{eqnarray}\nonumber
&&\left\langle T^{\beta}(t)\right\rangle = T_0^{\beta}e^{-rt}\left(1+\frac{t}{\tau_0}\right)^{-a\beta}+\\&& +T_0^{\beta}\tilde{r}e^{\tilde{r}}\left(E_{a\beta}(\tilde{r})-\left(1+\frac{t}{\tau_0}\right)^{1-a\beta}E_{a\beta}\left(\tilde{r}\left(1+\frac{t}{\tau_0}\right)\right)\right)\,.\nonumber\\
\end{eqnarray}
In the long-time limit $t\to\infty$
\begin{eqnarray}
\left\langle T^{\beta}\right\rangle = T_0^{\beta}\tilde{r}e^{\tilde{r}}E_{a\beta}(\tilde{r})
\end{eqnarray}
and the variance attains the form
\begin{equation}\label{sigmagen}
\sigma^2(T)=\left\langle T^{2}\right\rangle-\left\langle T\right\rangle^{2}=T_0^2\tilde{r}e^{\tilde{r}}\left(E_{2a}(\tilde{r})-\tilde{r}e^{\tilde{r}}E^2_ a(\tilde{r})\right)\,.
\end{equation}
%The variance is shown in Fig.~\ref{GDr} for $a=2,\,5/3$.
%\bigskip

\begin{figure}\centerline{\includegraphics[width=0.98\columnwidth]{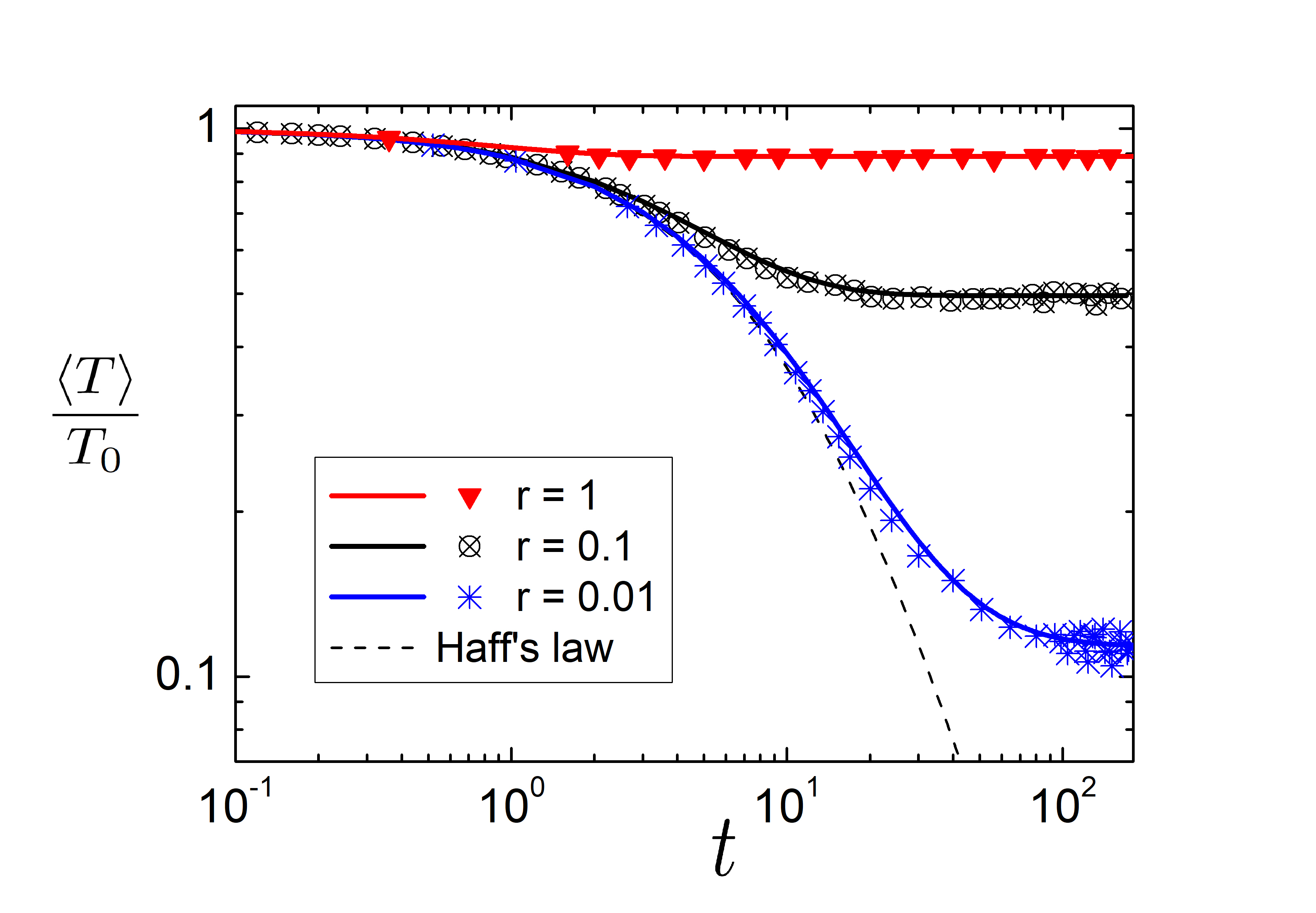}}\caption{Event-driven simulations of the time evolution of the granular temperature $T(t)$ in comparison with the theory. The characteristic time of the granular temperature decay $\tau_0=15.3$. Dashed line shows the evolution of a granular temperature of a force-free granular system without resetting according to the Haff's law. Lines correspond to analytical results (Eq.~\ref{Tmain}) and symbols correspond to results of event-driven simulations.} \label{GTresAwad}
\end{figure}

\subsection{Constant restitution coefficient}

The PDF (Eq.~\ref{ftv}) in the case of a constant restitution coefficient ($a=2$) becomes equal to
\begin{eqnarray}\label{fT}
&&f\left(T,t\right)= \delta\left(T-T_0\left(1+\frac{t}{\tau_0}\right)^{-2}\right)\!e^{-rt}+\\
&&\frac{\tilde{r}}{2T}\sqrt{\frac{T_0}{T}}e^{\tilde{r}\left(1-\sqrt{T_0/T}\right)}\Theta\left(T_0-T\right)\Theta\left(T-T_0\left(1+\frac{t}{\tau_0}\right)^{-2}\right)\,.\nonumber
\end{eqnarray}
The evolution of the PDF (Eq.~\ref{fT}) can also be obtained by simple MC simulations of a granular system with resetting. The simulation for $N=10^6$ different systems was performed, and the PDFs at different time moments are shown in Fig.~\ref{GfTmany}. The peak shifts towards smaller granular temperatures over time and eventually disappears.
The steady-state value of the PDF can be obtained in the long time limit $t\to\infty$
\begin{eqnarray}
f(T)= \frac{\tilde{r}}{2T}\sqrt{\frac{T_0}{T}}e^{\tilde{r}\left(1-\sqrt{T_0/T}\right)}\Theta\left(T_0-T\right)\Theta\left(T\right)\,.
\label{fT0}
\end{eqnarray}
This expression is depicted by the dashed line in Fig.~\ref{GfTmany}. 
%In the case of a constant restitution coefficient ($a=2$) 
The mean granular temperature can be derived from the PDF using Eq.~(\ref{Tmean}) or the expression
\begin{equation}\label{Tpsi}\left\langle T(t)\right\rangle =  T_0\left(1+\frac{t}{\tau_0}\right)^{-2}\!e^{-rt} + r\int\limits_0^t  T_0\left(1+\frac{\tau}{\tau_0}\right)^{-2}\!e^{-r\tau} d\tau\,.
\end{equation}
The direct integration of this equation yields
\begin{eqnarray}\nonumber
&&\left\langle T(t)\right\rangle = T_0\left[\left(1+\frac{t}{\tau_0}\right)^{-2}\!e^{-rt}+\tilde{r} - \tilde{r}e^{-r t}\left(1 + \frac{t}{\tau_0}\right)^{-1} \right.\\&&+\left.\,\tilde{r}^2 e^{\tilde{r}}\left(\rm{Ei}\left(-\tilde{\textit{r}}\right) -  \rm{Ei}\left(-\tilde{\textit{r}} \left(1 + \frac{\textit{t}}{\tau_0}\right)\right)\right)\right]\,,
\label{Tmain}\end{eqnarray}
%\begin{eqnarray}\nonumber
%&&\left\langle T(t)\right\rangle = T_0\left[\left(1+\frac{t}{\tau_0}\right)^{-2}\!e^{-rt}+r\tau_0 - r\tau_0e^{-r t}\left(1 + \frac{t}{\tau_0}\right)^{-1} \right.\\&&+\left.\,\left(r\tau_0\right)^2 e^{r\tau_0}\left(\rm{Ei}\left(-\textit{r}\tau_0\right) -  \rm{Ei}\left(-\textit{r}\tau_0 \left(1 + \frac{\textit{t}}{\tau_0}\right)\right)\right)\right]\,,
%\label{Tmain}\end{eqnarray}
where $\rm{Ei}(\textit{z})$ is exponential integral function \cite{abram}:
\begin{equation}\label{Ei}
\rm{Ei}(\textit{z})=-\int_{-\textit{z}}^{\infty}\frac{\textit{e}^{-\textit{t}}}{\textit{t}}\textit{dt}\,.
\end{equation}
Eq.~(\ref{Tmain}) may be also obtained by integrating Eq.~(\ref{Tmean}) by parts.
 %The comparison with event-driven simulations is shown in Fig.~\ref{GTresAwad}. A nice agreement is observed.
The constant value of the granular temperature has the form
\begin{eqnarray}\
T_c =\tilde{r}T_0\left(1-\tilde{r}\Gamma\left(0,\tilde{r}\right)e^{\tilde{r}}\right)\,,
\label{Teq}
\end{eqnarray}
where $\Gamma\left(0,\tilde{r}\right)$ is the incomplete gamma function. 
At $\tilde{r}\to 0$ the constant average granular temperature scales as
\begin{equation}\label{Tas}
T_c =  T_0 \left(\tilde{r}+\left(\gamma+\log \tilde{r}\right)\tilde{r}^2+o(\tilde{r}^2)\right)\,,
\end{equation}
where $\gamma$ is Euler's constant. At $\tilde{r}\to\infty$ Eq.~(\ref{Teq}) becomes
\begin{equation}\label{Tasbig}
T_c =  T_0 \left(1-\frac{2}{\tilde{r}}+\frac{6}{\tilde{r}^2}+o\left(\frac{1}{\tilde{r}^2}\right)\right)\,.
\end{equation}

The evolution of the mean granular temperature $\left\langle T(t)\right\rangle$ according to Eq.~(\ref{Tmain}) is shown in Fig. \ref{GTresAwad}. %is in good agreement with the event-driven simulation results (symbols in Fig. \ref{GTresAwad}). 
At the beginning of the evolution, it decreases according to Haff's law and then reaches a constant value $T_{c}$:
%The same value may be obtained in MC simulations by averaging both over the ensemble and time $t\gg 1/r$. 
At low resetting rates $r\to 0$ the mean time between the resetting events tends to infinity and the granular temperature evolves freely according to Haff's law (Eq.~\ref{Thaff}). If the resetting events are very frequent, and the resetting rate is much larger then the cooling rate, $r\gg \tau_0^{-1}$, the temperature remains constant, close to its initial value, $T\simeq T_0$.

\begin{figure}\centerline{\includegraphics[width=0.98\columnwidth]{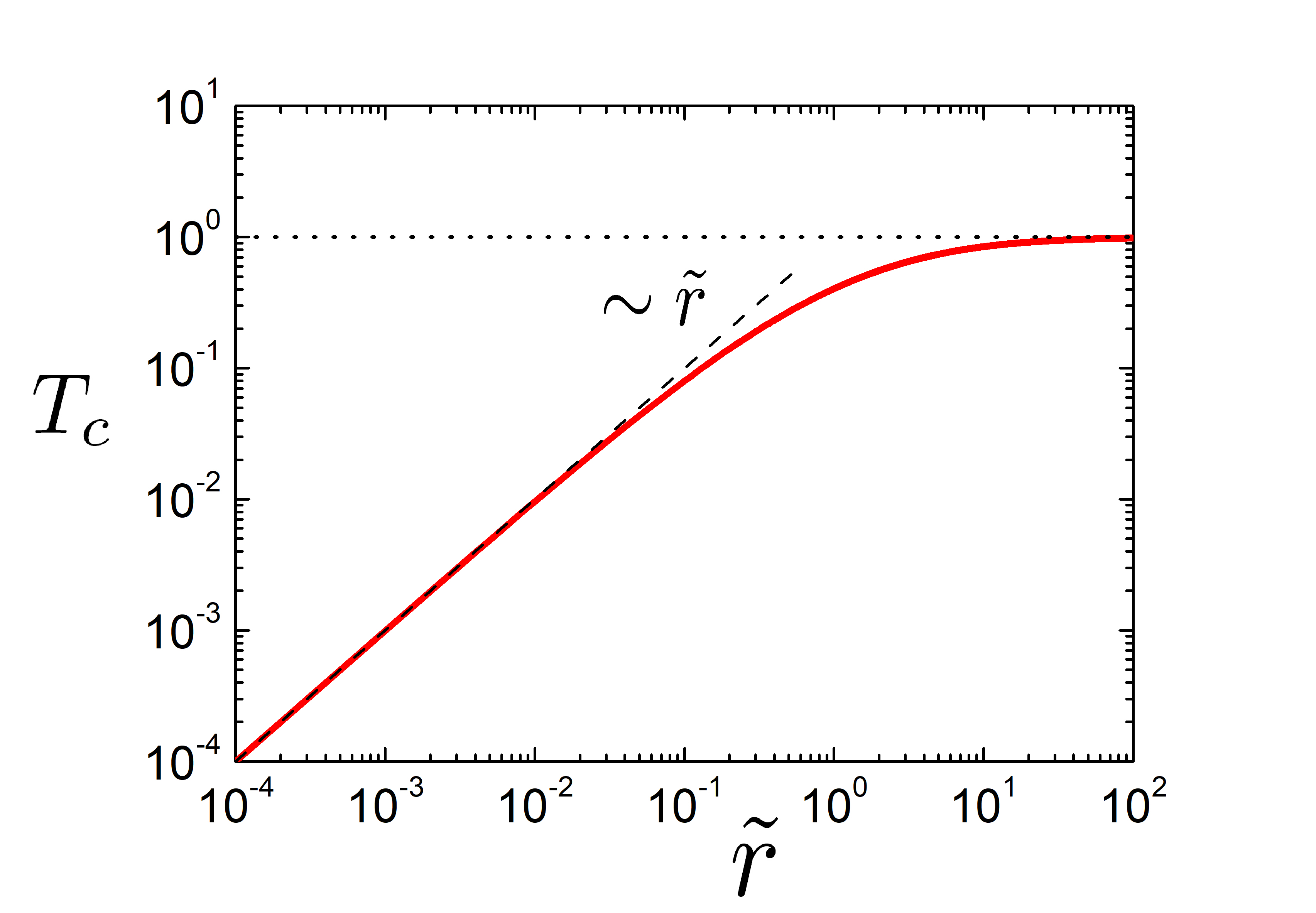}}\caption{Dependence of the steady-state value of granular temperature $T_c$ on the ratio between resetting and cooling rates $\tilde{r}=2r/\xi(0)$, Eq.~(\ref{Tgen}). Thick red line corresponds to a granular gas of particles, colliding with a constant restitution coefficient.
The dashed line shows linear dependence at short times, $T_c\sim \tilde{r}$. At large $\tilde{r}$ the average granular temperature tends to the initial value $T_0=1$, depicted with a dotted line %(Eq.~\ref{Tasbig})
.} \label{GTr}\end{figure}

The dependence of steady-state granular temperature $T_c$ on $\tilde{r}$ is shown in Fig.~\ref{GTr}. At low resetting rates compared with the cooling rates, $\tilde{r}\ll 1$, it increases linearly according to Eq.~(\ref{Tas}), and at high reduced resetting rates $\tilde{r}\gg 1$, it remains close to the initial value $T_0=1$. In the latter case the temperature was essentially reduced between the subsequent resetting events.

\begin{figure}\centerline{\includegraphics[width=0.98\columnwidth]{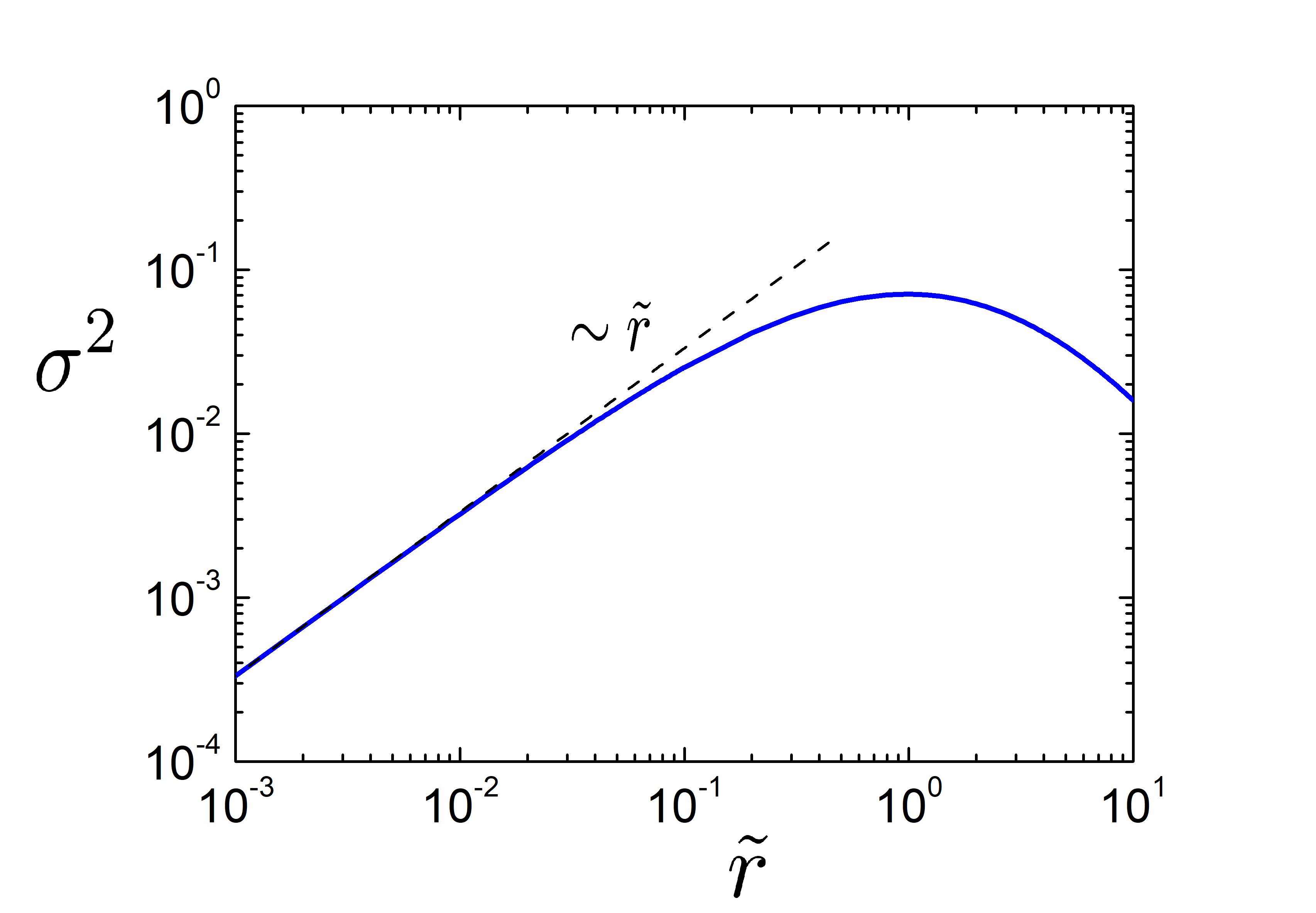}}\caption{Dependence of variance of granular temperature $\sigma^2(T)$ on the ratio between resetting and cooling rates $\tilde{r}=2r/\xi(0)$, Eq.~(\ref{sigmac}).  The dashed line shows linear dependence at short times, $\sigma^2(T)\sim \tilde{r}$.} \label{GDr}\end{figure}

The variance of the granular temperature in the steady state for granular gas with a constant restitution coefficient may be written according to Eq.~(\ref{sigmagen}):
\begin{equation}\label{sigmac}
\sigma^2(T)=T_0^2\tilde{r}e^{\tilde{r}}\left(E_{4}(\tilde{r})-\tilde{r}e^{\tilde{r}}E^2_2(\tilde{r})\right)\,.
\end{equation}
The dependence of the variance of granular temperatures on the reduced resetting rate $\tilde{r}$ is shown in Fig.~\ref{GDr}. At $\tilde{r}\to 0$ the variance scales according to
\begin{equation}\label{Das}\sigma^2(T)=T_0^2\left(\frac{1}{3}\tilde{r}-\frac{7}{6}\tilde{r}^2+o(\tilde{r}^2)\right)\,.\end{equation}
At low resetting rates, the variance increases linearly, as depicted by the dashed line in Fig.~\ref{GDr}. Subsequently, it reaches its maximum value and then decreases. This has a clear physical explanation: at very low and very high resetting rates, the value of the granular temperature is strictly determined. In the former case, it decreases according to Haff's law, whereas in the latter case, it tends to a constant value. The most diverse behavior appears in the system at an intermediate resetting rate $\tilde{r}\simeq 1$.

\section{Event-driven simulations}

We perform event-driven MD simulations \cite{compbook} of a dilute force free three dimensional granular gas of smooth particles colliding with a constant restitution coefficient $\epsilon$. We consider identical grains modeled as hard spheres of equal mass ($m=1$) and diameter ($\sigma=1$).
	In our simulations we undertake a system of $N_{\rm par}=10^5$ particles in a three-dimensional cubic box  with edge $L=158$ corresponding to a number density $n\approx0.025$. This number density is small and validates the assumption of binary collisions. The particles move freely between two successive collisions. After each collision the velocities of the particles are updated according to the collision rules given by Eqs.~(\ref{v1v2}).
		
	The simulation starts by randomly placing the particles in the box with periodic boundary conditions. The particles are assigned velocity vectors of equal magnitudes but in random directions. We ensure that the initial total momentum $\sum_{i} m{\bf v}_{i}(0)$ was zero with high accuracy. Initially, we let the system relax to the Maxwellian velocity distribution by allowing particles to collide elastically ($\varepsilon=1$) . This takes place in a few collisions per particle resulting into a system of homogeneous granular gas with Maxwellian velocity distribution and we consider this configuration (positions and velocities) as system's initial condition. Thereafter, we assign $\varepsilon=0.9$ and let the system evolve with time via inelastic collisions. We confirm that the system follows Haff's law and remains in a homogeneous cooling state (HCS) throughout the simulation. We have taken the initial granular temperature $T(0)=400/3$ and have obtained the results by averaging over $800$ independent initial conditions. The initial mean collision time (Eq.~\ref{tauc}) is equal to $\tau_c(0)=0.485$, while the  
    granular temperature relaxation time (Eq.~\ref{tau02}) significantly exceeds this value, $\tau_0=15.3=31.6\tau_c(0)$.

 \begin{figure}\centerline{\includegraphics[width=0.98\columnwidth]{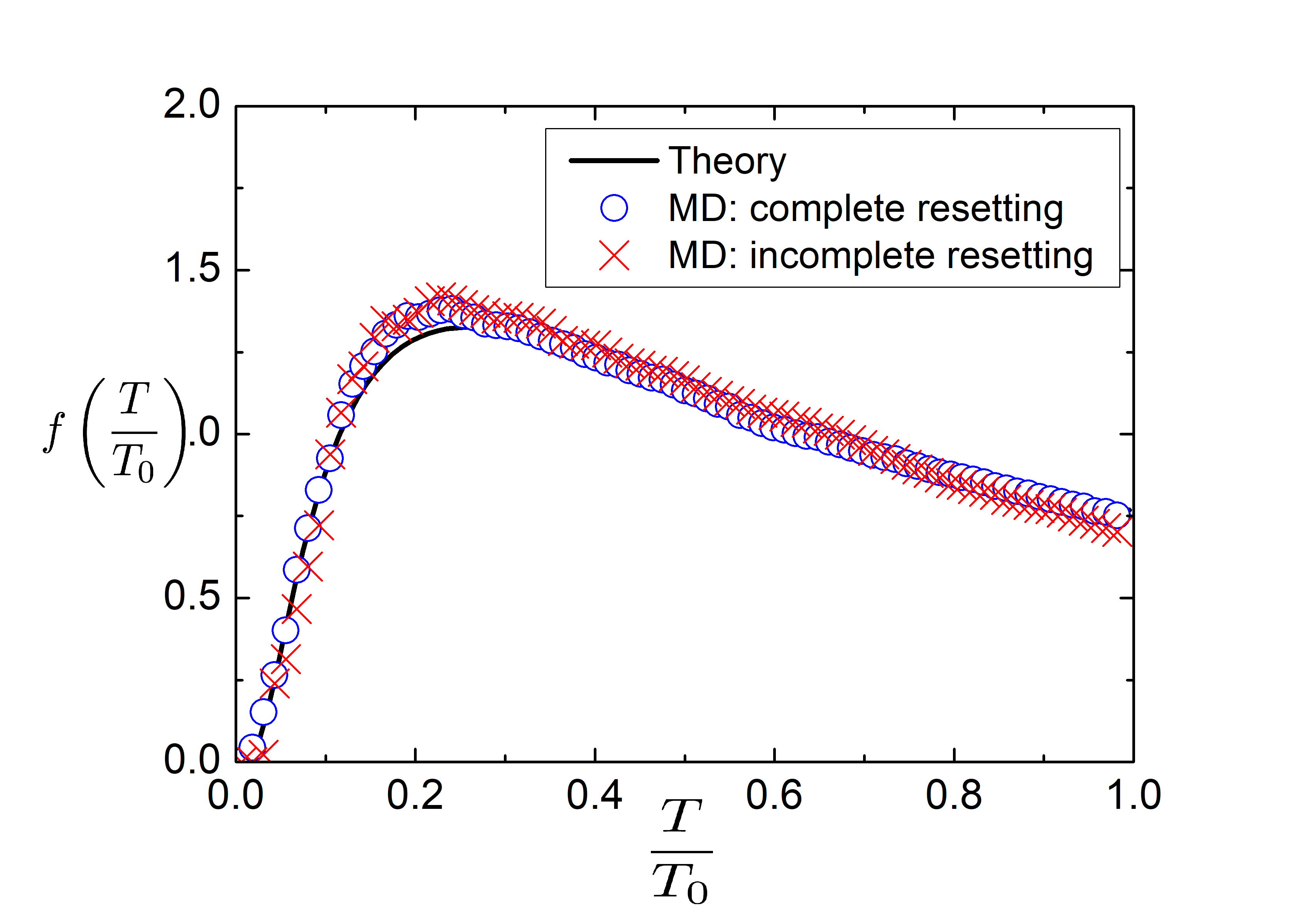}}\caption{Probability density function of granular temperatures $f(T)$ for $\tau_0=15.3$,  $r=0.1$, $t=100$. Symbols denote the results of event-driven simulations for both complete and incomplete resetting, line corresponds to the analytical form of the steady-state distribution of granular temperatures (Eq.~\ref{fT0}).
} \label{GfTAwad}
\end{figure}

 	The resetting occurs by assigning initial velocities to all the constituent particles at randomly selected times for which the waiting time distribution is given by Eq.~(\ref{pdfexp}). The system regains its initial temperature $T(0)=400/3$ after resetting. 
%The typical evolution of a granular system with a resetting rate $r=2$ is shown in Fig.~\ref{Gr2}.
The evolution of the mean granular temperature $\left\langle T(t)\right\rangle$, obtained in the simulations, is in good agreement with the theory, Eq.~(\ref{Tmain}) (Fig. \ref{GTresAwad}). A comparison of the granular temperature PDF, obtained in the simulations with the theoretical value is shown in Fig.~\ref{GfTAwad} and excellent agreement is again observed. 

Blue circles in Fig.~\ref{GfTAwad} correspond to the case of complete resetting, when all velocities of the particles are updated. In addition, we consider the case of the incomplete resetting (shown as red crosses in Fig.~\ref{GfTAwad}). We randomly choose half of the particles and update their velocities such that the total kinetic energy (or granular temperature) regains its initial value. Thus, if the velocity of any $i^{th}$ particle from the randomly selected ones is $\bf{v}_{i}$ then it is updated according to 	
	 \begin{eqnarray}
	 \textbf{v}_{i}^{s}= \sqrt{\frac{T(0)-T_{r}(t)}{T_c(t)}}\textbf{v}_{i} 	%\bf{v}_{i}^{s}=\alpha\bf{v}_{i} 
	 	 \end{eqnarray}
	Here, $T_c(t)$ is the temperature of the chosen half of the particles and $T_r(t)$ is the temperature of the rest half. The comparison in Fig.~\ref{GfTAwad} shows that the incompleteness of resetting does not affect essentially the steady-state distribution of granular temperature. 
    Both the characteristic relaxation time $\tau_0$ (Eq.~\ref{tau02}) of the granular temperature evolution and the mean time between the resetting events $r^{-1}$ significantly exceed the mean collision time $\tau_c$ in our system. The non-Maxwellian velocity distribution which arises during the incomplete resetting relaxes much faster compared to the resetting and granular gas cooling time scales. Therefore, the evolution of the system between the incomplete resetting events may be also described in terms of the Haff's law (Eq.~\ref{Thaff}). In such a way our theory is applicable for this particular type of incomplete resetting and also may be valid for a variety of different resetting (heating) protocols.
    %\red{This phenomenon appears because the magnitude and evolution of the granular temperature between resetting events do not significantly depend on the particular type of heating.}
    %In such a way, our theory may be valid for different resetting protocols.}

\section{Conclusions}

Granular systems with resets were investigated. We assume that at the resetting event, the granular temperature is instantly reset to the initial value. This may occur when the container with the granular material is shaken from time to time, a magnetic force is applied, or if energy is instantly %supplied 
inserted into the granular system in a different manner. The mean granular temperature was derived analytically and calculated in terms of both event-driven and Monte Carlo simulations. After a certain relaxation time it tends to a constant value, which decreases with increasing the average interval between resetting events. %The ensemble average granular temperature coincides with time average granular temperature for one system in the long time limit. 
The variance in the granular temperature reaches its maximum at intermediate resetting rates, tending to zero for both vanishing and extremely large resetting rates. Thus, the granular temperature of the driven granular medium may be controlled by the resetting rate. Our results may be helpful for understanding non-periodically driven granular systems.

\section{Acknowledgement} A.S.B. thanks Erez Aghion for the fruitful discussions. A.V.C. acknowledges support of the BMBF under the PLASMA-SPIN-ENERGY project. A.K.D. gratefully acknowledges the support and the resources provided by ‘PARAM Shivay Facility’ under the National Supercomputing Mission, Government of India at the Indian Institute of Technology, Varanasi.

\end{document}